\documentclass[prb,showpacs,amsmath,amssymb,preprint]{revtex4}

\usepackage{graphicx}
\usepackage{dcolumn}
\usepackage{bm}

\begin{document}

\title{Scattering of elastic waves by elastic spheres in a NaCl-type phononic crystal}
\author{Huanyang Chen, Xudong Luo \footnote{Corresponding author. Email address:
luoxd@sjtu.edu.cn} and Hongru Ma}

\affiliation{Institute of Theoretical Physics, Shanghai Jiao Tong
University, Shanghai 200240, China}%

\begin{abstract}
Based on the formalism developed by Psarobas et al [Phys. Rev. B
62, 278(2000)], which using the multiple scattering theory to
calculate properties of simple phononic crystals, we propose a
very simple method to study the NaCl-type phononic crystal. The
NaCl-type phononic crystal consists of two kinds of
non-overlapping elastic spheres with different mass densities,
$L\acute{a}me$ coefficients and radius following the same
periodicity of the ions in the real NaCl crystal. We focus on the
(001) surface, and view the crystal as a sequence of planes of
spheres, each plane of spheres has identical 2D periodicity. We
obtained the complex band structure of the infinite crystal
associated with this plane, and also calculated the transmission,
reflection and absorption coefficients for an elastic wave
(longitudinal or transverse) incident, at any angle, on a slab of
the crystal of finite thickness.

\end{abstract}


\maketitle

\section{Introduction}

Recently, inspired by the remarkable properties of the photonic
crystals\cite{y}, there has  been increasing interest in the study
of phononic crystals\cite{es,khm,mjt,lzmz,skek}, which are
constructed by composing identical particles (the inclusions) on
host medium periodically. The acoustic behaviors in such materials
are different from those in ordinary materials, such as the
frequency band structure of classical waves and the propagation of
elastic waves in the materials. Various applications are also
proposed, for example, acoustic lens\cite{cssm},
waveguides\cite{mi} and negative refraction materials\cite{sss}.
The behaviors of the new materials are related to their band
structures so that the calculation of the acoustic wave spectrum,
or band structure is the key point in the studies of phononic
crystals.

There are several techniques for the calculation of frequency band
structures, the plane-wave approach is the simplest though the
convergence is slow\cite{es,khm,mjt}, the multiple-scattering
theory (MST) or KKR (Korringa, Kohn, and Rostoker) approach are
widely used with better convergence\cite{psm,lcsgp,sspm}, and the
finite difference time domain (FDTD) method is a new and promising
method\cite{t}. When the inclusions of phononic crystals are
spheres in 3-dimensional case or cylinders in 2-dimensional case,
layer MST is a very effective calculation technique not only for
the evaluation of band structures, but also for the calculation of
the transmission, reflection and absorption coefficients of a slab
of the crystal of finite thickness. Based on the layer MST
approach, Psarobas et al \cite{psm,sspm} have developed a
formalism for calculating the frequency band structure of phononic
crystals and the corresponding scattering properties. In their
algorithm and the related  program package, they consider only the
3-dimensional phononic crystals
 formed by  one inclusion per unit cell, so the phononic crystals
are the so called ``simple'' crystals. In addition to band structure, their
program package can also deal with scattering problems with a slab of
crystals of finite thickness, if all successional scattering planes
are 2-dimensional ``simple'' lattices.

However, it is desirable to have a technique to deal with
``complex'' phononic crystals (i.e. crystals have more than one
type of inclusion per unit cell). The crystals with ``complex''
structures may offer us different band gaps and new properties
which are not existed in  ``simple'' crystals. By generalizing the
simple lattice layer MST method, we propose here a simple
technique to deal with NaCl-type ``complex'' phononic crystals by
means of MST. Calculation of the band structures is made and
compared to the ``simple'' case. A  general method to work with
``complex'' phononic crystals, which are similar to those in low
energy electron diffraction of ``complex'' crystals\cite{p}, is
also discussed in Appendix A.  Our method is based on MST and
different from the plane-wave approach suggested by Zhang et
al\cite{zllw}, so it is more efficient in dealing with spherical
inclusion, and the case of fluid-solid composites. In addition,
one may notice that the NaCl-type crystal can also be seen as a
succession of two nonoverlapping planes of different spheres in
the (111) direction and considered as a heterostructure
\cite{sspm} or as a case of a super cell with planar defects
\cite{psm2}, so the complex band structure could be extracted from
their program. Actually, Liu et al\cite{lcs} have also used their
bulk KKR method to calculate the full band structure of some
complex phononic crystals with HCP and Diamond structure. However,
the layer KKR method here we used is necessary, since it can also
be used to calculate the transmission, reflection and absorption
coefficients for an elastic wave incident on a slab of NaCl-type
``complex'' crystal. Recently, it has also been applied to a
chiral structure, which is regarded as a more complex
heterostructure, many interesting results was found \cite{cc}. The
programm package developed by Sainidou et al\cite{sspm} is a
stable and efficient package in treating the simple phononic
crystals, we have used the package extensively in our ``complex''
phononic crystal calculation.

The paper is organized as follows. In section \ref{theory}, we present a simple
method to evaluate the scattering properties of a slab consisting a
number of layers, which could contain two kinds of spheres with
specific 2D periodic structure. In section \ref{results}, we demonstrate the
applicability of our method with some numerical results and
discussions. Finally, the conclusions of the paper are given in Sec. \ref{conc}.

\section{Theory} \label{theory}
In this section, we briefly review the theory of layer MST used in
the calculation of band structures of phononic crystals with the
NaCl-type structure. Firstly, we consider a ``complex'' 2D crystal
in $xy$-plane at $z=0$. There are two kinds of spheres in the 2D
crystal. The first is denoted as type A, located on the sites of a
2D lattice specified by
\begin{equation}\label{eq14}
\mathbf{R}_{n}=n_{1}\mathbf{a}_{1}+n_{2}\mathbf{a}_{2},
\end{equation}
where $\mathbf{a}_{1}$,$\mathbf{a}_{2}$ are the primitive vectors
of a 2D lattice in the $xy$-plane, and
$n_{1},n_{2}=0,\pm1,\pm2,\pm3,...$.  The second kind of spheres is
denoted as type B, located at
\begin{equation}\label{eq15}
\mathbf{R}_{n}^{b}=\overrightarrow{ab}+\mathbf{R}_{n}
  =\overrightarrow{ab}+n_{1}\mathbf{a}_{1}+n_{2}\mathbf{a}_{2},
\end{equation}
where $\overrightarrow{ab}=(\mathbf{a}_{1}+\mathbf{a}_{2})/2$ for
NaCl-type phononic crystal. In this case, there are two types of
inclusion per unit cell, and the corresponding 2D reciprocal
lattice is
\begin{equation}\label{eq16}
\mathbf{g}=m_{1}\mathbf{b}_{1}+m_{2}\mathbf{b}_{2},
\end{equation}
where $m_{1},m_{2}=0,\pm1,\pm2,\pm3,...$ and vectors
$\mathbf{b}_{1}$, $\mathbf{b}_{2}$ are defined by
\begin{equation}\label{eq17}
\mathbf{b}_{i}\cdot\mathbf{a}_{j}=2\pi\delta_{ij},  \quad\quad
i,j=1,2.
\end{equation}

Now we consider that a plane wave, either longitudinal or
transverse, is incident onto such a plane of spheres of type A and
B. The displacement vector of the incident wave has the
form\cite{psm},
\begin{equation}\label{eq18}
\mathbf{u}_{in}^{s'}(\mathbf{r})=
\sum_{i'}[u_{in}]_{\mathbf{g}'i'}^{s'}
\exp(i\mathbf{K}_{\mathbf{g}'\nu'}^{s'}
\cdot\mathbf{r})\hat{\mathbf{e}}_{i'},
\end{equation}
where $s'=+(-)$ represents that a wave is incident onto the plane
from the left(right), $\nu'$ denotes the polarization of the
incident wave: $q_{\nu'}=q_{l}=\omega / c_l$ for a longitudinal
wave and $q_{\nu'}=q_{t}=\omega / c_t$ for a transverse one.
Because of the 2D periodicity in the plane of spheres, wavevector
can be rewritten as
\begin{equation}\label{eq19}
\mathbf{K}_{\mathbf{g}'\nu'}^{\pm}\equiv\mathbf{k}_{||}
+\mathbf{g}'\pm[q_{\nu'}^{2}-(\mathbf{k}_{||}+\mathbf{g}')^2]^{1/2}
\hat{\mathbf{e}}_{z},
\end{equation}
where $\hat{\mathbf{e}}_{z}$ is the unit vector along the $z$
axis, $\mathbf{k}_{||}$ is the reduced wave vector lying in the
surface Brillouin zone (SBZ) of the given lattice, and
$\mathbf{g}'$ is one of the reciprocal lattice vectors in
Eq.(\ref{eq16}).

In order to avoid the cumbersome formulas in the representation of
the displacement field, we use the same complete set of
spherical-wave solutions as recommended by Sainidou et
al\cite{sspm},
\begin{equation}\begin{array}{ll}\label{eq13}
\mathbf{J}_{Llm}(\mathbf{r}) =
\frac{1}{q_{l}}\nabla[j_{l}(q_{l}r)Y_{l}^{m}(\hat{\mathbf{r}})], &
\quad \mathbf{H}_{Llm}(\mathbf{r}) =
\frac{1}{q_{l}}\nabla[h_{l}^{+}(q_{l}r)Y_{l}^{m}(\hat{\mathbf{r}})],\\
\mathbf{J}_{Mlm}(\mathbf{r}) =
j_{l}(q_{t}r)\mathbf{X}_{lm}(\hat{\mathbf{r}}), & \quad
\mathbf{H}_{Mlm}(\mathbf{r}) =
h_{l}^{+}(q_{t}r)\mathbf{X}_{lm}(\hat{\mathbf{r}}),\\
\mathbf{J}_{Nlm}(\mathbf{r}) = \frac{i}{q_{t}}\nabla\times
j_{l}(q_{t}r)\mathbf{X}_{lm}(\hat{\mathbf{r}}), & \quad
\mathbf{H}_{Nlm}(\mathbf{r}) = \frac{i}{q_{t}}\nabla\times
h_{l}^{+}(q_{t}r)\mathbf{X}_{lm}(\hat{\mathbf{r}}),
\end{array}\end{equation}
here $j_l$ and $h^+_l$ are the spherical Bessel and Hankel
functions, $Y_{l}^{m}$ and $\mathbf{X}_{lm}$ are the usual and
vector spherical harmonics, respectively.

The scattered waves by all spheres in the plane can be separated
into two parts. One is the contributions from spheres of type A,
$\mathbf{u}_{sc}^{As}(\mathbf{r})$, and the other is those from
spheres of type B, $\mathbf{u}_{sc}^{Bs}(\mathbf{r}')$, here
$\mathbf{r}=\mathbf{r}'+\overrightarrow{ab}$.  Let $a_{n,lm}^{+P}
(P=L,M,N)$ be the scatter wave coefficients of sphere A at the
site $R_n$. According to the Bloch theorem, we have
\begin{equation}\label{eq20}
\mathbf{u}_{sc}^{As}(\mathbf{r})
=\sum_{n,Plm}a_{n,lm}^{+P}\mathbf{H}_{Plm}(\mathbf{r}_{n})
=\sum_{Plm}a_{lm}^{+P}\sum_{\mathbf{R}_{n}}
\exp(i\mathbf{k}_{||}\cdot\mathbf{R}_{n})
\mathbf{H}_{Plm}(\mathbf{r}-\mathbf{R}_{n}).
\end{equation}
where $a_{lm}^{+P} (P=L,M,N)$ are the scatter wave coefficients of
sphere A at the origin. In the domain close to surface of sphere
of type B, it is always satisfied that
$|\mathbf{r}'|<|\mathbf{R}_{n}-\overrightarrow{ab}|$ since all
spheres are not overlapped. So that the
$\mathbf{u}_{sc}^{As}(\mathbf{r})$ can be expanded into spherical
waves about the sphere of type B, which is located at
$\overrightarrow{ab}$, as follows
\begin{equation}\label{eq21}
\mathbf{u}_{sc}^{As}(\mathbf{r})
=\exp(i\mathbf{K}_{\mathbf{g}'\nu'}^{s'}\cdot\overrightarrow{ab})
\sum_{Plm}d_{lm}^{'P}\mathbf{J}_{Plm}(\mathbf{r}').
\end{equation}
Likewise, we have
\begin{equation}\begin{array}{rcl}\label{eq22}
\mathbf{u}_{sc}^{Bs}(\mathbf{r}') &=& \displaystyle
\exp(i\mathbf{K}_{\mathbf{g}'\nu'}^{s'} \cdot\overrightarrow{ab})
\sum_{n,Plm}b_{n,lm}^{+P}\mathbf{H}_{Plm}(\mathbf{r}_{n}') \\
 &=& \displaystyle \exp(i\mathbf{K}_{\mathbf{g}'\nu'}^{s'}
\cdot\overrightarrow{ab})\sum_{Plm}b_{lm}^{+P}
\sum_{\mathbf{R}_{n}} \exp(i\mathbf{k}_{||}
\cdot\mathbf{R}_{n})\mathbf{H}_{Plm}(\mathbf{r}'-\mathbf{R}_{n}),
\end{array}\end{equation}
it can also be expanded as follows
\begin{equation}\label{eq23}
\mathbf{u}_{sc}^{Bs}(\mathbf{r}')
=\sum_{Plm}c_{lm}^{'P}\mathbf{J}_{Plm}(\mathbf{r}).
\end{equation}
In addition, the external incident wave is
\begin{equation}\label{eq30}
\mathbf{u}_{in}(\mathbf{r})
=\sum_{Plm}a_{lm}^{0P}\mathbf{J}_{Plm}(\mathbf{r})
=\exp(i\mathbf{K}_{\mathbf{g}'\nu'}^{s'}\cdot\overrightarrow{ab})
\sum_{Plm}a_{lm}^{0P}\mathbf{J}_{Plm}(\mathbf{r}').
\end{equation}

Moreover, if we remove the term corresponding to sphere located at
$\mathbf{R}_{n}=0$ in Eq. (\ref{eq20}), the terms left are just
the scattering waves come from all spheres of type A except for
itself. We denote it by $\mathbf{u}_{sc}^{As\prime}(\mathbf{r})$,
and expand it into spherical waves about the origin
\begin{equation}\label{eq24}
\mathbf{u}_{sc}^{As\prime}(\mathbf{r})
=\sum_{Plm}a_{lm}^{'P}\mathbf{J}_{Plm}(\mathbf{r}).
\end{equation}
In the same way, after removing the term corresponding to
$\mathbf{R}_{n}=0$ in Eq. (\ref{eq22}), we have
\begin{equation}\label{eq25}
\mathbf{u}_{sc}^{Bs\prime}(\mathbf{r}')
=\exp(i\mathbf{K}_{\mathbf{g}'\nu'}^{s'}\cdot\overrightarrow{ab})
\sum_{Plm}b_{lm}^{'P}\mathbf{J}_{Plm}(\mathbf{r}').
\end{equation}
It means the scattering waves come from all spheres of type B
except for the one located at $\overrightarrow{ab}$. It can be
shown from direct calculation (see Appendix A for the details)
that
\begin{equation}\begin{array}{cc}\label{eq29}
\displaystyle
a_{lm}^{'P}=\sum_{P'l'm'}\Omega_{lm;l'm'}^{PP'}a_{l'm'}^{+P'},
\quad & \quad \displaystyle
b_{lm}^{'P}=\sum_{P'l'm'}\Omega_{lm;l'm'}^{PP'}b_{l'm'}^{+P'}, \\
\displaystyle
c_{lm}^{'P}=\sum_{P'l'm'}\Xi_{lm;l'm'}^{PP'}b_{l'm'}^{+P'}, \quad
& \quad \displaystyle
d_{lm}^{'P}=\sum_{P'l'm'}\Xi_{lm;l'm'}^{PP'}a_{l'm'}^{+P'}.
\end{array}\end{equation}

According to MST, the wave incident on the sphere of type A
located at the origin consists of three parts, the first party is
the externally incident wave, the second part is the sum of all
the waves scattered by spheres of type A except for itself, and
the third part is the sum of all the waves scattered by spheres of
type B. The coefficients of the incident wave and scattering wave
is related to each other by the Mie matrix, so we have the
following relationship
\begin{equation}\label{eq31}
a_{lm}^{+P} =\sum_{P'l'm'}T_{a;lm;l'm'}^{PP'}(a_{l'm'}^{0P'}
+a_{l'm'}^{'P'}+c_{l'm'}^{'P'}).
\end{equation}
Likewise, the wave incident on the sphere B located at
$\overrightarrow{ab}$ also consists of three parts, the externally
incident wave, the sum of all the waves scattered by spheres of
type B except for itself, and the sum of all the waves scattered
by spheres of type A, so we have
\begin{equation}\label{eq32}
b_{lm}^{+P} =\sum_{P'l'm'}T_{b;lm;l'm'}^{PP'}\left(a_{l'm'}^{0P'}
+b_{l'm'}^{'P'}+d_{l'm'}^{'P'}\right).
\end{equation}
Combining Eqs. (\ref{eq29}), (\ref{eq31}) and (\ref{eq32}), we
obtain
\begin{equation}\begin{array}{l}\label{eq34}
\displaystyle
\sum_{P'l'm'}\left[\delta_{PP'}\delta_{ll'}\delta_{mm'}
-\sum_{P''l''m''}T_{a;lm;l''m''}^{PP''}\Omega_{l''m'';l'm'}^{P''P'}\right]
a_{l'm'}^{+P'} \\
\qquad\qquad\qquad\quad \displaystyle
+\sum_{P'l'm'}\left[-\sum_{P''l''m''}T_{a;lm;l''m''}^{PP''}
\Xi_{l''m'';l'm'}^{P''P'}\right]b_{l'm'}^{+P'}
  = \sum_{P'l'm'}T_{a;lm;l'm'}^{PP'}a_{l'm'}^{0P'}, \\
\displaystyle
\sum_{P'l'm'}\left[-\sum_{P''l''m''}T_{b;lm;l''m''}^{PP''}
\Xi_{l''m'';l'm'}^{P''P'}\right] a_{l'm'}^{+P'} \\
\quad\quad \displaystyle
+\sum_{P'l'm'}\left[\delta_{PP'}\delta_{ll'}\delta_{mm'}
-\sum_{P''l''m''}T_{b;lm;l''m''}^{PP''}\Omega_{l''m'';l'm'}^{P''P'}\right]
b_{l'm'}^{+P'}  = \sum_{P'l'm'}T_{b;lm;l'm'}^{PP'}a_{l'm'}^{0P'}.
\end{array}\end{equation}
When the coefficients $a_{lm}^{0P}$ of the incident wave are
given, equations (\ref{eq34}) determine the scattering
coefficients $a_{lm}^{+P}$ and $b_{lm}^{+P}$ of spheres of type A
and B. Consequently, the waves scattered from the plane of spheres
of type A and B are also obtained in terms of Eqs. (\ref{eq20})
and (\ref{eq22}), respectively.

We write the coefficients $a_{lm}^{0P}$ in the form of
\begin{equation}\label{eq35}
a_{lm}^{0P}=\sum_{i'}A_{lm;i'}^{0P}
(\mathbf{K}_{\mathbf{g}'\nu'}^{s'})[u_{in}]_{\mathbf{g}'i'}^{s'}
\end{equation}
based on Eq. (\ref{eq18}), where coefficients $A_{lm}^{0P}$ are
given in Psarobas et al\cite{psm} (Eqs. (3.4), (3.7) and (3.8)).
Because of the linearity of Eqs. (\ref{eq34}), the coefficients
$a_{lm}^{+P}$ and $b_{lm}^{+P}$ can also be written as follows
\begin{equation}\label{eq36}
a_{lm}^{+P}=\sum_{i'}A_{lm;i'}^{+P}(\mathbf{K}_{\mathbf{g}'\nu'}^{s'})
[u_{in}]_{\mathbf{g}'i'}^{s'},
\end{equation}
\begin{equation}\label{eq37}
b_{lm}^{+P}=\sum_{i'}B_{lm;i'}^{+P}(\mathbf{K}_{\mathbf{g}'\nu'}^{s'})
[u_{in}]_{\mathbf{g}'i'}^{s'},
\end{equation}
so that the Eqs. (\ref{eq34}) are reduced to
\begin{equation}\begin{array}{l}\label{eq38}
\displaystyle
\sum_{P'l'm'}\left[\delta_{PP'}\delta_{ll'}\delta_{mm'}
-\sum_{P''l''m''}T_{a;lm;l''m''}^{PP''}
\Omega_{l''m'';l'm'}^{P''P'}\right]
A_{l'm';i'}^{+P'}(\mathbf{K}_{\mathbf{g}'\nu'}^{s'}) \\
\displaystyle \quad
+\sum_{P'l'm'}\left[-\sum_{P''l''m''}T_{a;lm;l''m''}^{PP''}
\Xi_{l''m'';l'm'}^{P''P'}\right]
B_{l'm';i'}^{+P'}(\mathbf{K}_{\mathbf{g}'\nu'}^{s'})
=\sum_{P'l'm'}T_{a;lm;l'm'}^{PP'}
A_{l'm';i'}^{0P'}(\mathbf{K}_{\mathbf{g}'\nu'}^{s'}), \\
\displaystyle
\sum_{P'l'm'}\left[-\sum_{P''l''m''}T_{b;lm;l''m''}^{PP''}
\Xi_{l''m'';l'm'}^{P''P'}\right]
A_{l'm';i'}^{+P'}(\mathbf{K}_{\mathbf{g}'\nu'}^{s'})  \\
\displaystyle \quad
+\sum_{P'l'm'}\left[\delta_{PP'}\delta_{ll'}\delta_{mm'}
-\sum_{P''l''m''}T_{b;lm;l''m''}^{PP''}\Omega_{l''m'';l'm'}^{P''P'}\right]
B_{l'm';i'}^{+P'}(\mathbf{K}_{\mathbf{g}'\nu'}^{s'}) \\
\displaystyle \quad =\sum_{P'l'm'}T_{b;lm;l'm'}^{PP'}
A_{l'm';i'}^{0P'}(\mathbf{K}_{\mathbf{g}'\nu'}^{s'}).
\end{array}\end{equation}
Now, the scattered waves given by Eqs. (\ref{eq20}) and
(\ref{eq22}) are expressed as sum of plane waves as follows (see
Appendix B)
\begin{equation}\begin{array}{rcl}\label{eq40}
\mathbf{u}_{sc}^{As}(\mathbf{r})&=& \displaystyle
\sum_{\mathbf{g}i}[u_{sc}^{A}]_{\mathbf{g}i}^{s}
\exp(i\mathbf{K}_{\mathbf{g}\nu}^{s}\cdot\mathbf{r})\hat{\mathbf{e}}_{i}, \\
\mathbf{u}_{sc}^{Bs}(\mathbf{r}')&=& \displaystyle
\exp(i\mathbf{K}_{\mathbf{g}'\nu'}^{s'}\cdot\overrightarrow{ab})
\sum_{\mathbf{g}i}[u_{sc}^{\prime B}]_{\mathbf{g}i}^{s}
\exp(i\mathbf{K}_{\mathbf{g}\nu}^{s}\cdot\mathbf{r}')\hat{\mathbf{e}}_{i}
=\sum_{\mathbf{g}i}[u_{sc}^{B}]_{\mathbf{g}i}^{s}
\exp(i\mathbf{K}_{\mathbf{g}\nu}^{s}\cdot\mathbf{r})\hat{\mathbf{e}}_{i}.
\end{array}\end{equation}
Here $u_{sc}^{\prime B}$ is obtained in coordinates ${\bf
r}^\prime$, it is obviously that its expression is similar to
$u_{sc}^{A}$. The total scattered wave then becomes
\begin{equation}\label{eq42}
\mathbf{u}_{sc}^{s}(\mathbf{r})=
\sum_{\mathbf{g}i}([u_{sc}^{A}]_{\mathbf{g}i}^{s}
+[u_{sc}^{B}]_{\mathbf{g}i}^{s})
\exp(i\mathbf{K}_{\mathbf{g}\nu}^{s}\cdot\mathbf{r})\hat{\mathbf{e}}_{i},
\end{equation}
where $[u_{sc}^{A}]_{\mathbf{g}i}^{s}$ and
$[u_{sc}^{B}]_{\mathbf{g}i}^{s}$ are given by
\begin{equation}\label{eq43}
[u_{sc}^{A}]_{\mathbf{g}i}^{s}=
\sum_{i'}\sum_{Plm}\Delta_{lm;i}^{P}(\mathbf{K}_{\mathbf{g}\nu}^{s})
A_{lm;i'}^{+P}(\mathbf{K}_{\mathbf{g}'\nu'}^{s'})[u_{in}]_{\mathbf{g}'i'}^{s'},
\end{equation}
\begin{equation}\label{eq44}
[u_{sc}^{B}]_{\mathbf{g}i}^{s}=
\sum_{i'}\sum_{Plm}\Delta_{lm;i}^{P}(\mathbf{K}_{\mathbf{g}\nu}^{s})
B_{lm;i'}^{+P}(\mathbf{K}_{\mathbf{g}'\nu'}^{s'})[u_{in}]_{\mathbf{g}'i'}^{s'}
\exp{(i(\mathbf{K}_{\mathbf{g}'\nu'}^{s'}-\mathbf{K}_{\mathbf{g}\nu}^{s})
\cdot \overrightarrow{ab})} .
\end{equation}
The derivation is given in Appendix B.

With these results, we get this 2D ``complex'' crystal's
$\mathbf{M}$-matrix elements
\begin{equation}\label{eq45}
M_{\mathbf{g}i;\mathbf{g}'i'}^{ss'}=
  \delta_{ss'}\delta_{\mathbf{g}\mathbf{g}'}\delta_{ii'}
  +\sum_{Plm}\Delta_{lm;i}^{P}(\mathbf{K}_{\mathbf{g}\nu}^{s})
  [A_{lm;i'}^{+P}(\mathbf{K}_{\mathbf{g}'\nu'}^{s'})
  +B_{lm;i'}^{+P}(\mathbf{K}_{\mathbf{g}'\nu'}^{s'})
  \exp{(i(\mathbf{K}_{\mathbf{g}'\nu'}^{s'}-\mathbf{K}_{\mathbf{g}\nu}^{s})
\cdot \overrightarrow{ab})}].
\end{equation}
Using the algorithm recommended by Psarobas et al\cite{psm,sspm},
as soon as the $\mathbf{M}$-matrix is given, both the frequency
band structure of an infinite crystal and acoustic properties of a
slab of this crystal can be calculated in the same way.  Here we
only write down the $\mathbf{Q}$-matrix elements
\begin{equation}\begin{array}{rcl}\label{eq46}
Q_{\mathbf{g}i;\mathbf{g}'i'}^{I} &=&
M_{\mathbf{g}i;\mathbf{g}'i'}^{++}
  \exp\left[i(\mathbf{K}_{\mathbf{g}\nu}^{+}\cdot\mathbf{d}_{r}
  +\mathbf{K}_{\mathbf{g}'\nu'}^{+}\cdot\mathbf{d}_{l})\right], \\
Q_{\mathbf{g}i;\mathbf{g}'i'}^{II} &=&
M_{\mathbf{g}i;\mathbf{g}'i'}^{+-}
  \exp\left[i(\mathbf{K}_{\mathbf{g}\nu}^{+}\cdot\mathbf{d}_{r}
  -\mathbf{K}_{\mathbf{g}'\nu'}^{-}\cdot\mathbf{d}_{r})\right], \\
Q_{\mathbf{g}i;\mathbf{g}'i'}^{III} &=&
M_{\mathbf{g}i;\mathbf{g}'i'}^{-+}
  \exp\left[-i(\mathbf{K}_{\mathbf{g}\nu}^{-}\cdot\mathbf{d}_{l}
  -\mathbf{K}_{\mathbf{g}'\nu'}^{+}\cdot\mathbf{d}_{l})\right], \\
Q_{\mathbf{g}i;\mathbf{g}'i'}^{IV} &=&
M_{\mathbf{g}i;\mathbf{g}'i'}^{--}
  \exp\left[-i(\mathbf{K}_{\mathbf{g}\nu}^{-}\cdot\mathbf{d}_{l}
  +\mathbf{K}_{\mathbf{g}'\nu'}^{-}\cdot\mathbf{d}_{r})\right].
\end{array}\end{equation}

\section{Numerical results and discussion} \label{results}
We demonstrate the applicability of our method by applying it to two
specific examples. Firstly, we illustrate the 2D periodic structure
of NaCl-type phononic crystal in figure 1, here the nearest distance
between the same spheres is defined as lattice constant $a_{0}$, so
that $\mathbf{a}_{1}=a_{0}(1,0,0)$ and $\mathbf{a}_{2}=a_{0}(0,1,0)$
in Eqs. (\ref{eq14}) and (\ref{eq15}). Then we choose
$\mathbf{d}_{l}=\mathbf{d}_{r}=a_{0}(0.25,0.25,\sqrt{2}/4)$ in Eq.
(\ref{eq46})\cite{sspm}, and obtain a slab of NaCl-type phononic
crystal of finite thickness associated with the (001) surface. We
take $a_{0}=1cm$ in the calculation.

\begin{figure}[ht]
\includegraphics{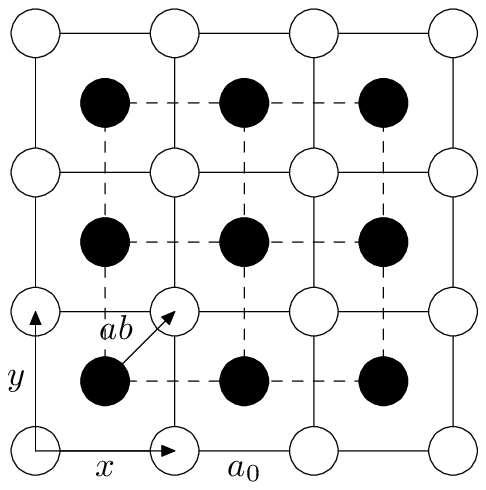}\hspace{3.5cm}
\includegraphics{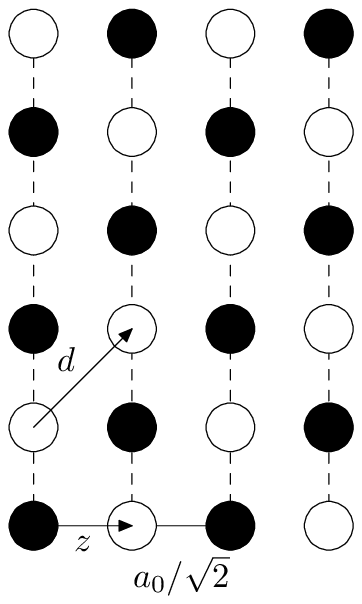} \caption{The 2D
structure of each layer of (001) plane is given in the left graph,
the lattice constant $a_{0}$ is the nearest distance between the
same spheres, the distance between two neighbor layers is
$a_{0}/\sqrt{2}$ for NaCl-type phononic crystal (the right graph
is the section perpendicular to the layer plane).}
        \label{fig1}
\end{figure}

In figure 2, we show the transmittance curves of a slab of 8
layers parallel to the (001) surface of NaCl-type phononic crystal
at normal incident for longitudinal wave (a) and transverse wave
(b). The materials are: sphere of type A is lead sphere of radius
$S=0.25 cm$, sphere of type B is lead sphere of radius $S=0.10
cm$, the host matrix is epoxy. The relevant parameters are, for
lead: $\rho=11.6\times 10^{3} kg/m^{3}$, $c_{l}=2490 m/s$,
$c_{t}=1130 m/s$, and for epoxy: $\rho=1.18\times 10^{3}
kg/m^{3}$, $c_{l}=2540 m/s$, $c_{t}=1160 m/s$. Figure 3 shows the
corresponding complex band structure of the infinite NaCl-type
phononic crystal associated with the (001) surface. All the units
are following the suggestion given by Sainidou et al~\cite{sspm},
where $c_{0}=2540 m/s$.

\begin{figure}[ht]
    \centering
        \begin{minipage}[t]{.46\textwidth}
            \resizebox{6.5cm}{!}{\includegraphics*{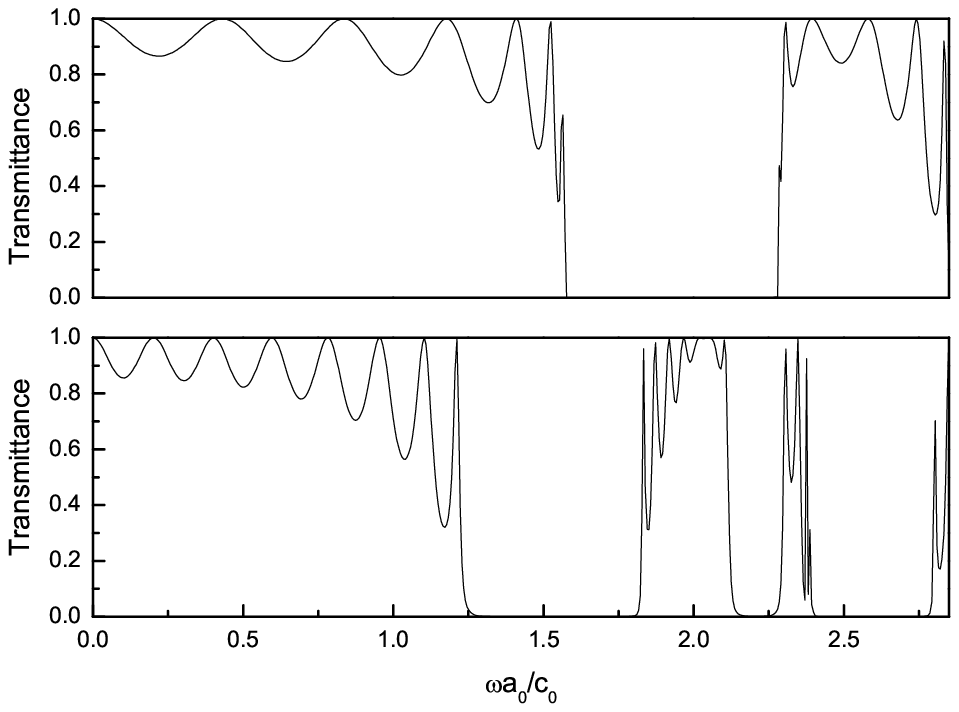}}
            \caption{The transmittance curves of a slab of 8 layers parallel to
(001) surface of NaCl-type phononic crystal at normal incident for
longitudinal wave (a the upper graph) and transverse wave (b the
lower graph). }
            \label{fig2}
        \end{minipage}
        \hfill
             \begin{minipage}[t]{0.46\textwidth}
            \resizebox{6.5cm}{!}{\includegraphics*{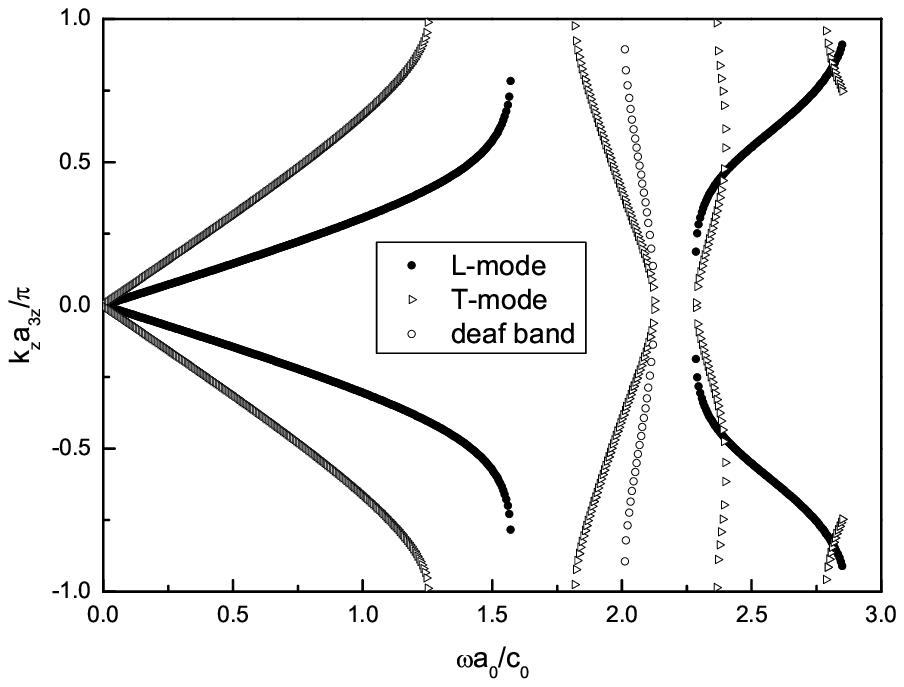}}
            \caption{The phononic band structure at the center of the SBZ of a
(001) surface of NaCl-type phononic crystal corresponding to
Fig.2. The filled circles represent the longitudinal mode, the
open triangles represent the transverse modes and the open circles
represent the deaf band, respectively.}
            \label{fig3}
        \end{minipage}
\end{figure}

Comparing the results in figure 2 and figure 3, we see that there
is a frequency band gap approximate between  1.6 to 2.25 for the
longitudinal wave, and also several band gaps for the transverse
wave. It should be noted that the width of the gaps and the
positions have all changed from those of the traditional
``simple'' phononic crystal, so that a lot of work has to be done
for searching some new properties of this kind of phononic
crystals.

In figure 4 we show the reflectance curve of a slab of layers
parallel to the same surface at normal incident for longitudinal
wave. The materials are taken as follows: sphere of type A is lead
sphere of radius $S=0.20 cm$ coated with a $0.5 mm$ layer of
silicone rubber, sphere of type B is silica sphere of radius
$S=0.10cm$ coated with a $1.5 mm$ layer of silicone rubber, the
host matrix is epoxy. The relevant parameters are, for lead:
$\rho=11.6\times10^{3}  kg/m^{3}$, $c_{l}=2490 m/s$, $c_{t}=1130
m/s$, for silicone rubber: $\rho=1.3\times10^{3}kg/m^{3}$,
$c_{l}=22.87 m/s$, $c_{t}=5.55 m/s$, for silica: $\rho=2.2\times
10^{3}kg/m^{3}$, $c_{l}=5970 m/s$, $c_{t}=3760 m/s$, and for
epoxy: $\rho=1.18\times 10^{3} kg/m^{3}$, $c_{l}=2540 m/s$,
$c_{t}=1160 m/s$. We take such materials that the spheres of type
A and B have different local resonance frequencies, which can be
estimated in the approach of Liu et al\cite{lzmz}.

When the NaCl-type phononic crystal is consisted of two kinds of
local resonance materials, our results show that both of local the
resonance frequency (see the peaks with amplitude close to 1 that
represents total reflectance) are still keep unchanged, which
means that this kind of hybrid does not change the property of
intrinsic local resonance. In addition, as the number of the
layers increased, the band gap is broadened indicate by the
broader peak. We also see that there is a small wiggle near
$0.11$. In fact, when we scan the frequency in more refined steps,
we found that there is a very sharp resonance located at there.
The corresponding frequency band structure is shown in figure 5.
\begin{figure}[ht]
    \centering
        \begin{minipage}[t]{.46\textwidth}
            \resizebox{6.5cm}{!}{\includegraphics*{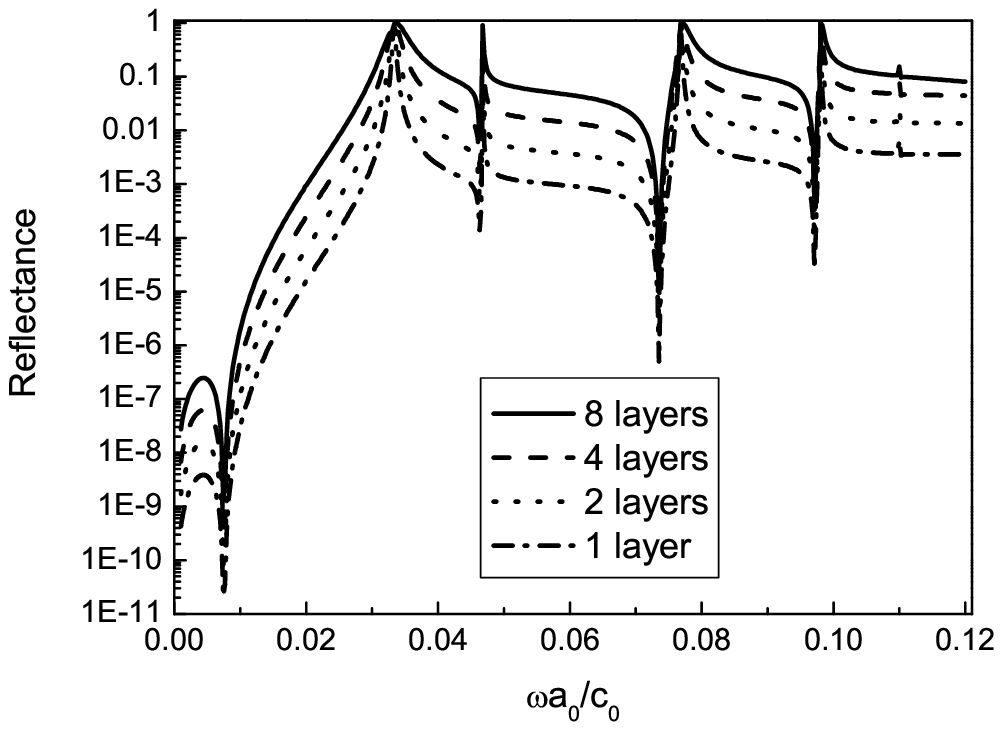}}
            \caption{The reflectance curves of a slab of different layers
parallel to (001) surface of NaCl-type phononic crystal at normal
incident for longitudinal wave for local resonance materials. The
solid, dash, dot and dot-dash line are for 8, 4, 2 and single
layer, respectively.}
            \label{fig4}
        \end{minipage}
        \hfill
             \begin{minipage}[t]{0.46\textwidth}
            \resizebox{6.5cm}{!}{\includegraphics*{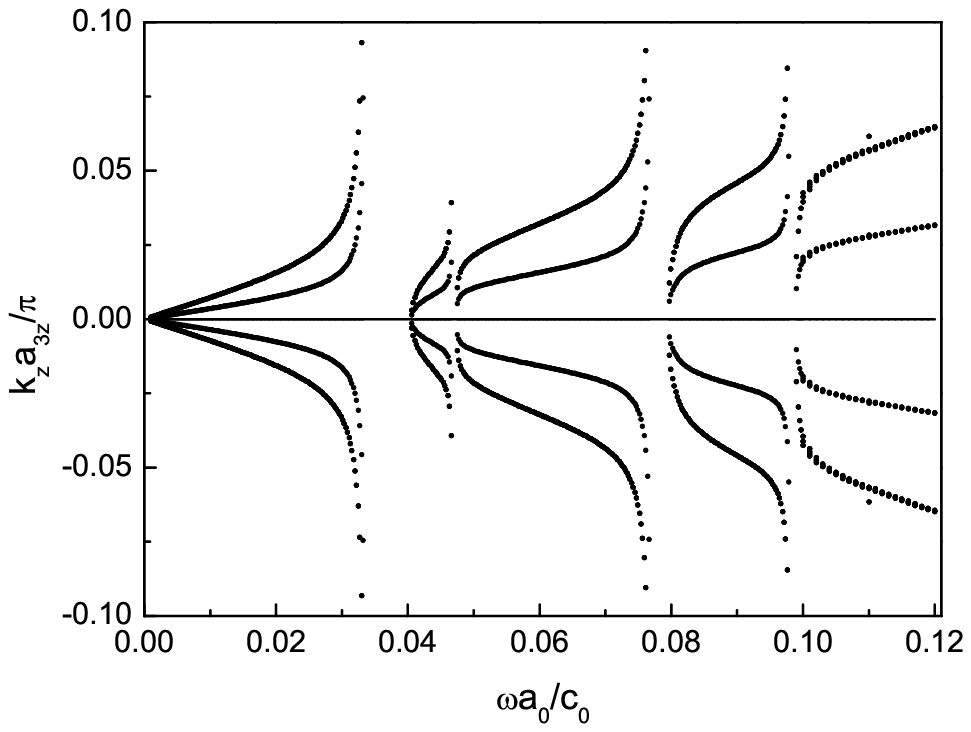}}
            \caption{The phononic band structure at the center of the SBZ of
a (001) surface of NaCl-type phononic crystal corresponding to
Fig.4.}
            \label{fig5}
        \end{minipage}
\end{figure}

The examples here are concerning the normal incident wave for
simplicity, one can also take any incidence angles at will. Our
results obtained with an angular momentum cutoff $l_{max}=4$ and
$13$ $\mathbf{g}$ vectors, the relative error of the results are
about $10^{-3}$. To obtain the error estimates, we have also
performed the calculation with larger cutoffs of angular momentum
and compared the results with different cutoffs. In addition, we
have also compared our results of identical spheres, which becomes
a ``simple'' crystal in fact, to the results of the known programm
of ``simple'' phononic crystal for confirming our code.

\section{Conclusion} \label{conc}
We have shown that, for a system of two kinds of non-overlapping
elastic spheres with different mass density, $L\acute{a}me$
coefficients and radius forming a  NaCl-type structure, one can
calculate the phononic band structure of the infinite crystal
accurately and efficiently, and one can also obtain the
transmission, reflection, and absorption coefficients of elastic
waves incident onto a slab of the materials of finite thickness
using the method and formalism of the present paper.

The programm used in this work is a direct extension of the
programm package by Sainidou et al\cite{sspm}, we thank the
authors of the package for providing this useful package. The work
is supported by the National Natural Science Foundation of China
under grant No.90103035, No.10334020 and No.10174041.

\appendix

\section{}
In this appendix, we give the detail derivation of equation
(\ref{eq29}). Since Eq. (\ref{eq20}) and Eq. (\ref{eq21}) are
exactly the same, we have
\begin{equation}
\sum_{Plm}d_{lm}^{'P}\mathbf{J}_{Plm}(\mathbf{r}')=
   \exp(-i(\mathbf{k}_{||}+\mathbf{g}')\cdot\overrightarrow{ab})
     \sum_{Plm}a_{lm}^{+P}\sum_{\mathbf{R}_{n}}
     \exp(i\mathbf{k}_{||}\cdot\mathbf{R}_{n})
     \mathbf{H}_{Plm}(\mathbf{r}'+\overrightarrow{ab}-\mathbf{R}_{n}),
\end{equation}
using the fact that $\exp(i\mathbf{g}'\cdot\mathbf{R}_{n})=1$, it
becomes
\begin{equation}
\sum_{Plm}d_{lm}^{'P}\mathbf{J}_{Plm}(\mathbf{r}')=
\sum_{Plm}a_{lm}^{+P}\sum_{\mathbf{R}_{n}}
 \exp(i(\mathbf{k}_{||}+\mathbf{g}')\cdot(\mathbf{R}_{n}-\overrightarrow{ab}))
 \mathbf{H}_{Plm}(\mathbf{r}'-(\mathbf{R}_{n}-\overrightarrow{ab})).
\end{equation}
The vector spherical waves originated at
$\overrightarrow{ab}-\mathbf{R}_{n}$ and those originated at
$\overrightarrow{ab}$ are related by the following expression (for
example, see Eq. (B9) in Appedix B of Sainidou et al\cite{ssm},
whose matrix $\Omega$ is just the matrix $\Lambda$ here.),
\begin{equation}
\mathbf{H}_{Plm}(\mathbf{r}'-(\mathbf{R}_{n}-\overrightarrow{ab}))
=\sum_{P'l'm'}
\Lambda_{lm;l'm'}^{PP'}(-(\mathbf{R}_{n}-\overrightarrow{ab}))
\mathbf{J}_{P'l'm'}(\mathbf{r}'),
\end{equation}
so that
\begin{equation}\label{eqa4}
\Xi_{lm;l'm'}^{PP'}=\sum_{\mathbf{R}_{n}}
\exp[i(\mathbf{k}_{||}+\mathbf{g}')\cdot(\mathbf{R}_{n}-\overrightarrow{ab})]
\;\Lambda_{lm;l'm'}^{PP'}(-(\mathbf{R}_{n}-\overrightarrow{ab})).
\end{equation}
Moreover, in the case of ``simple'' crystal, Sanididou et
al\cite{sspm} remarked that the evaluation of the matrices
$\Omega_{lm;l'm'}^{PP'}$ becomes the evaluation of a well-known
matrix $Z$ in the theory of low-energy electron
diffraction\cite{p}. Here the matrices $\Xi_{lm;l'm'}^{PP'}$ can
also evaluated in the same way except that the definition of
matrix $Z$ is modified as follows
\begin{equation}\label{eqa5}
Z_{lm}^{l'm'}(q_{t}) = \sum_{\mathbf{R}_{n}}
\exp[i(\mathbf{k}_{||}+\mathbf{g}')\cdot(\mathbf{R}_{n}-\overrightarrow{ab})]
\; G_{lm;l'm'}(-(\mathbf{R}_{n}-\overrightarrow{ab});q_{t}),
\end{equation}
where
\begin{equation}
G_{lm;l''m''}(-\mathbf{\tau};q_{t})
=\sum_{l'm'}E_{lm}(l'm';l''m'')h_{l'}^{+}(q_{t}\tau)
Y_{l'}^{-m'}(-\hat{\mathbf{\tau}}),
\end{equation}
\begin{equation}
E_{lm}(l'm';l''m'')=4\pi(-1)^{(l-l'-l'')/2}(-1)^{m'+m''}\int
d\hat{r}Y_{l}^{m}(\hat{r})Y_{l'}^{m'}(\hat{r})Y_{l''}^{-m''}(\hat{r}).
\end{equation}

In order to evaluate matrix $Z$ in eq. (\ref{eqa5}), we introduce
matrix K whose elements are
\begin{equation}
K_{lm}^{l'm'}(q_{t})=\sum_{\mathbf{R}_{n}\neq 0}
\exp[i(\mathbf{k}_{||}+\mathbf{g}')\cdot\mathbf{R}_{n}] \;
G_{lm;l'm'}(-\mathbf{R}_{n};q_{t}),
\end{equation}
which can be numerically calculated by using Kambe's
method\cite{k,p}. Based on the following relationship\cite{ssm}
\begin{equation}
h_{l'}^{+}(q_{t}|\mathbf{R}_{n} -\overrightarrow{ab}|)Y_{l'}^{-m'}
(-\mathbf{R}_{n}+\overrightarrow{ab}) =\sum_{l'''m'''}
\xi_{l'-m';l'''-m'''}
(\overrightarrow{ab};q_{t})h_{l'''}^{+}(q_{t}R_{n})Y_{l'''}^{-m'''}
(-\mathbf{R}_{n}),
\end{equation}
where
\begin{equation}
\xi_{lm;l''m''}(-\mathbf{\tau};q_{t})
=\sum_{l'm'}E_{lm}(l'm';l''m'')j_{l'}(q_{t}\tau)
Y_{l'}^{-m'}(-\hat{\mathbf{\tau}}),
\end{equation}
and some straightforward but trivial algebra calculation, we
finally get
\begin{equation}\begin{array}{rcl}
Z_{lm}^{l''m''}(q_{t}) &=& \displaystyle
\exp[-i(\mathbf{k}_{||}+\mathbf{g}')\cdot\overrightarrow{ab}]
\sum_{l'm'}\{ E_{lm}(l'm';l''m'')  \\
&& \displaystyle \quad \quad  \times
[h_{l'}^{+}(q_{t}ab)Y_{l'}^{-m'} (\overrightarrow{ab})
+\sum_{LM}K_{LM}^{l'm'}(q_{t})j_{L}(q_{t}ab)
Y_{L}^{-M}(\overrightarrow{ab}) ]\}.
\end{array}\end{equation}
After that, we obtain the matrix Z through matrix K, so are matrix
$\Xi$. Similarly we get another matrix $\Xi$ in Eq.(\ref{eq29}).
In addition, it should be remarked that those two matrix $\Xi$ are
different in general, except for a few cases, such as NaCl-type.
For the two matrices $\Omega$ in Eq. (\ref{eq29}), we have
\begin{equation}\label{eqa12}
\Omega_{lm;l'm'}^{PP'}=\sum_{\mathbf{R}_{n}\neq\mathbf{0}}
\exp(i\mathbf{k}_{||}\cdot\mathbf{R}_{n})
\Lambda_{lm;l'm'}^{PP'}(-\mathbf{R}_{n}).
\end{equation}
These matrices also appear in the case of ``simple'' crystal, its
evaluation is already available\cite{sspm}.

However, there are some tricks for several specific examples to
calculate the $\Xi$ matrix in eq. (\ref{eqa4}). Here we propose
the NaCl-type for instance. Supposing the spheres of type A is
identical with those of type B, the crystal becomes a ``simple''
crystal and forms a new 2D periodic structure, its lattice is
denoted by $\mathbf{R}_{n}'$ now. It is quite easy to obtain that
\begin{equation}\label{eqa13}
\Xi_{lm;l'm'}^{PP'}=\sum_{\mathbf{R}_{n}'\neq\mathbf{0}}
\exp[i(\mathbf{k}_{||}+\mathbf{g}') \cdot \mathbf{R}_{n}']
\Lambda_{lm;l'm'}^{PP'}(-\mathbf{R}_{n}')
-\sum_{\mathbf{R}_{n}\neq\mathbf{0}} \exp[i
(\mathbf{k}_{||}+\mathbf{g}') \cdot \mathbf{R}_{n}]
\Lambda_{lm;l'm'}^{PP'}(-\mathbf{R}_{n}).
\end{equation}
Based on the computer programm of ``simple'' crystal, it is
obviously that only little modification is needed for dealing with
the NaCl-type crystal. This is actually our original idea, there
are also other types can use this idea, for example, when
$\overrightarrow{ab}=\mathbf{a}_{1}/2$ or
$\overrightarrow{ab}=\mathbf{a}_{2}/2$.

The specific form of the above matrix elements are given by
Sainidou et al\cite{sspm}. It should be remarked that $\Xi$ have
the same property of $\Omega$ if the spheres of type A and type B
locate in the same plane, it says
\begin{equation}\begin{array}{rl}
\Xi_{lm;l'm'}^{MM}=\Xi_{lm;l'm'}^{NN}=\Xi_{lm;l'm'}^{LL}=0, &
\quad \hbox{unless} \;\; l+m+l'+m': \;\hbox{even} , \\
\Xi_{lm;l'm'}^{MN}=\Xi_{lm;l'm'}^{NM}=0, & \quad \hbox{unless}
\;\; l+m+l'+m': \;\hbox{odd}.
\end{array}\end{equation}

\section{}
Firstly, we write down an identity\cite{psm},
\begin{equation}
\sum_{\mathbf{R}_{n}}\exp(i\mathbf{k}_{||}\cdot\mathbf{R}_{n})
h_{l}^{+}(q_{\nu}r_{n})Y_{l}^{m}(\hat{\mathbf{r}}_{n})
=\sum_{\mathbf{g}}\frac{2\pi(-i)^{l}}{q_{\nu}A_{0}K_{\mathbf{g}\nu
z}^{+}}Y_{l}^{m}(\hat{\mathbf{K}}_{\mathbf{g}\nu}^{\pm})
\exp(i\mathbf{K}_{\mathbf{g}\nu}^{\pm}\cdot\mathbf{r}),
\end{equation}
where $A_{0}$ denotes the area of the unit cell of the lattice
given by Eq. (\ref{eq14}), the plus (minus) sign on
$\mathbf{K}_{\mathbf{g}\nu}$ must be used for $z>0$ ($z<0$).

Using the identity above, we write Eq. (\ref{eq20}) as follows,
\begin{equation}\label{eqb2}
\mathbf{u}_{sc}^{As}(\mathbf{r})
=\sum_{\mathbf{g}}\sum_{Plm}a_{lm}^{+P}
\Delta_{lm}^{P}(\mathbf{K}_{\mathbf{g}\nu}^{s})
\exp(i\mathbf{K}_{\mathbf{g}\nu}^{s}\cdot\mathbf{r}),
\end{equation}
where
\begin{equation}\begin{array}{rcl}
\Delta_{lm}^{L}(\mathbf{K}_{\mathbf{g}l}^{s}) &=&\displaystyle
\frac{2\pi(-i)^{l-1}}{q_{l}A_{0}K_{\mathbf{g}l
z}^{+}}Y_{l}^{m}(\hat{\mathbf{K}}_{\mathbf{g}l}^{s})
\hat{\mathbf{e}}_{1}, \\
\Delta_{lm}^{M}(\mathbf{K}_{\mathbf{g}t}^{s}) &=& \displaystyle
\frac{2\pi(-i)^{l}}{q_{t}A_{0}K_{\mathbf{g}t
z}^{+}\sqrt{l(l+1)}}\{ [\alpha_{l}^{-m}\cos\theta
e^{i\phi}Y_{l}^{m-1}(\hat{\mathbf{K}}_{\mathbf{g}t}^{s})
-m\sin\theta Y_{l}^{m}(\hat{\mathbf{K}}_{\mathbf{g}t}^{s})\\
& &\displaystyle +\alpha_{l}^{m}\cos\theta
e^{-i\phi}Y_{l}^{m+1}(\hat{\mathbf{K}}_{\mathbf{g}t}^{s})]
\hat{\mathbf{e}}_{2} +i[\alpha_{l}^{-m}e^{i\phi}Y_{l}^{m-1}
(\hat{\mathbf{K}}_{\mathbf{g}t}^{s})-\alpha_{l}^{m}e^{-i\phi}Y_{l}^{m+1}
(\hat{\mathbf{K}}_{\mathbf{g}t}^{s})]\hat{\mathbf{e}}_{3} \}, \\
\Delta_{lm}^{N}(\mathbf{K}_{\mathbf{g}t}^{s}) &=& \displaystyle
\frac{2\pi(-i)^{l}}{q_{t}A_{0}K_{\mathbf{g}t z}^{+}\sqrt{l(l+1)}}
\{i[\alpha_{l}^{-m}e^{i\phi}Y_{l}^{m-1}
(\hat{\mathbf{K}}_{\mathbf{g}t}^{s}) -\alpha_{l}^{m}
e^{-i\phi}Y_{l}^{m+1}(\hat{\mathbf{K}}_{\mathbf{g}t}^{s})]
\hat{\mathbf{e}}_{2} \\
 & & \displaystyle -[\alpha_{l}^{-m}\cos\theta
e^{i\phi}Y_{l}^{m-1}(\hat{\mathbf{K}}_{\mathbf{g}t}^{s})-m
\sin\theta Y_{l}^{m}(\hat{\mathbf{K}}_{\mathbf{g}t}^{s})
+\alpha_{l}^{m}\cos\theta
e^{-i\phi}Y_{l}^{m+1}(\hat{\mathbf{K}}_{\mathbf{g}t}^{s})]
\hat{\mathbf{e}}_{3}\}.
\end{array}
\end{equation}
where $\alpha_{l}^{m}=\sqrt{(l-m)(l+m+1)}/2$, and $(\theta,\phi)$
denoting the angular variables of $\mathbf{K}_{\mathbf{g}t}^{s}$.
Substituting $a_{lm}^{+P}$ from Eq. (\ref{eq36}) into Eq.
(\ref{eqb2}), we obtain
\begin{equation}
\mathbf{u}_{sc}^{As}(\mathbf{r})
=\sum_{\mathbf{g}i}[u_{sc}^{A}]_{\mathbf{g}i}^{s}
\exp(i\mathbf{K}_{\mathbf{g}\nu}^{s}\cdot\mathbf{r})
\hat{\mathbf{e}}_{i},
\end{equation}
where
\begin{equation}
[u_{sc}^{A}]_{\mathbf{g}i}^{s}
=\sum_{i'}\sum_{Plm}\Delta_{lm;i}^{P}(\mathbf{K}_{\mathbf{g}\nu}^{s})
A_{lm;i'}^{+P}(\mathbf{K}_{\mathbf{g}'\nu'}^{s'})[u_{in}]_{\mathbf{g}'i'}^{s'}
\end{equation}


\begin{thebibliography}{100}

\bibitem{y} E. Yablonovitch, Phys. Rev. Lett. {\bf 58}, 2059(1987).

\bibitem{es} E. N. Economou and M. Sigalas, J. Acoust. Soc. Am. {\bf 95}, 1734(1994);
M. Sigalas and E. N. Economou, Europhys. Lett. {\bf 36}, 241(1996);
A. D. Klironomos and E. N. Economou, Solid State Commun. {\bf 105},
327(1998).

\bibitem{khm} M. S. Kushwaha, P. Halevi, G. Martinez, L. Dobrzynski and
B. Djafari-Rouhani, Phys. Rev. B {\bf 49}, 2313(1994).

\bibitem{mjt} F. R. Montero de Espinosa, E. Jimenez, and M. Torres,
Phys. Rev. Lett. {\bf 80}, 1208(1998).

\bibitem{lzmz} Z.Y. Liu, X.X. Zhang, Y. Mao, Y.Y. Zhu, Z. Yang, C.T. Chan, and
P. Sheng, Science {\bf 289}, 1734(2000).

\bibitem{skek} M. Sigalas, M.S. Kushwaha, E.N. Economou, M. Kafesaki, I.E.
Psarobas, and W. Steurer, Z. Kristallogr. {\bf 220},
765-809(2005).

\bibitem{psm}  I. E. Psarobas, N. Stefanou and A. Modinos,
Phys. Rev. B {\bf 62}, 278(2000).

\bibitem{psm2}  I. E. Psarobas, N. Stefanou and A. Modinos,
Phys. Rev. B {\bf 62}, 5536(2000).

\bibitem{lcsgp} Z. Liu, C. T. Chan, P. Sheng, A. L. Goertzen, and J. H. Page, Phys. Rev.
B {\bf 62}, 2446(2000).

\bibitem{sspm} R. Sainidou, N. Stefanou, I. E. Psarobas, and A. Modinos,
Comput. Phys. Commun. {\bf 166}, 197(2005).

\bibitem{cssm} F. Cervera, L. Sanchis, J. V. $S\acute{a}chez$-$P\acute{e}rez$,
R. $Mart\acute{i}nez$-Sala, C. Rubio, F. Meseguer, C.
$L\acute{o}pez$, D. Caballero, and J. $S\acute{a}nchez$-Dehesa,
Phys. Rev. Lett. {\bf 88}, 023902(2002).

\bibitem{mi} T. Miyashita and C. Inoue, Jpn. J. Appl. Phys., Part 1 {\bf 40}, 3488(2001).

\bibitem{sss} R. A. Shelby, D. R. Smith, and S. Schultz, Science {\bf 292}, 77(2001).

\bibitem{t} A. Taflove, {\it The Finite Difference Time Domain
Method} (Artech, Boston, 1998).

\bibitem{zllw} X. Zhang, Z. Liu, Y. Liu and F. Wu, Phys. Lett.
A, {\bf 313}, 455(2003).

\bibitem{lcs} Z.Y. Liu, C.T. Chan and P.Sheng, Phys. Rev. B.{\bf 65},
165116(2002)

\bibitem{p} J.B. Pendry, {\it Low Energy Electron Diffraction} (Academic, London,
1974). A. Modinos, Field, Thermionic and Secondary Electron
Emission Spectroscopy (Plenum Press, New York, 1984).

\bibitem{ssm} R. Sainidou, N. Stefanou and A. Modinos, Phys. Rev. B.
{\bf 69}, 064301(2004).

\bibitem{k} K. Kambe, Z. Naturforsch., {\bf 23a}, 1280(1968).

\bibitem{cc} H.Y. Chen and C.T. Chan, unpublished results.

\end{thebibliography}
\end{document}